\documentclass[12pt]{article}
\usepackage{amssymb}
\def\beq{\begin{equation}}
\def\eeq{\end{equation}}
\usepackage[all,2cell,dvips]{xy}

\def\IR{\relax{\rm I\kern -.18em R}}

\begin{document}
\title{Full hamiltonian structure for a parametric coupled Korteweg-de Vries system}
\author{ \Large  A. Restuccia*, A. Sotomayor**}
\maketitle{\centerline{*Departamento de F\'{\i}sica}}
\maketitle{\centerline{Universidad de Antofagasta} }
\maketitle{\centerline{*Departamento de F\'{\i}sica}}
\maketitle{\centerline{Universidad Sim\'on Bol\'{\i}var}}
\maketitle{\centerline{**Departamento de Matem\'aticas}}
\maketitle{\centerline{Universidad de Antofagasta}}
\maketitle{\centerline{e-mail:arestu@usb.ve,
 adrian.sotomayor@uantof.cl}}

\begin{abstract} We obtain the full hamiltonian structure for a parametric coupled KdV system.
The coupled system arises from four different real basic
lagrangians. The associated hamiltonian functionals and the
corresponding Poisson structures follow from the geometry of a
constrained phase space by using the Dirac approach for
constrained systems. The overall algebraic structure for the
system is given in terms of two pencils of Poisson structures
with associated hamiltonians depending on the parameter of the
Poisson pencils. The algebraic construction we present admits the
most general space of observables related to the coupled system.
\end{abstract}


Pacs: 02.30.Jr , 02.30.Ik, 11.10.Ef.

\section{Introduction} Coupled Korteweg-de Vries (KdV) systems describes several physical interactions of interest.
Hirota and Satsuma \cite{Hirota} proposed a model that describes
interactions of two long waves with different dispersion
relations. Gear and Grimshaw \cite{Gear} considered a coupled KdV
system to describe linearly stable internal waves in a stratified
fluid.

More recently Lou, Tong, Hu and Tang \cite{Lou} proposed models
which may be used in the description of atmospheric and oceanic
phenomena. Coupled KdV systems were also analyzed in
\cite{Karasu,Sakovich,Casati}. An important area of interest of
high energy physics related to coupled systems is provided by the
supersymmetric extensions of KdV equations
\cite{Kupershmidt,Mathieu1,Mathieu2,Krivonos1,Delduc,Krivonos2,Popowicz}
and more generally by operator and Clifford valued extensions of
KdV equation \cite{Olver,Restuccia}.

In this work we consider a parametric coupled KdV system. For
some values of the parameter, $\lambda<0$, the system corresponds
to the complexification of KdV equation. For $\lambda=0$ the
system corresponds to one of the Hirota-Satsuma coupled KdV
systems, while for $\lambda>0$ the system is equivalent to two
decoupled KdV equations. We analyze the hamiltonian formulation
and the associated Poisson bracket structure of the system.
Although some properties of the complexification of KdV arise
directly from the analogous ones on the solutions of the KdV
equation there are new properties, in particular, the full
hamiltonian structure, which does not have an analogous on the
original real equation. In fact, the complexification approach
gives rise only to holomorphic observables on phase space. The
full hamiltonian structure of the complex system give rise to
self-adjoint hamiltonian functionals, whose hamiltonian flow are
the complex KdV equations, and it provides the full structure of
observables on phase space, not only the holomorphic ones.

The approach we will follow in our analysis is to construct a
family of lagrangians from which the coupled KdV system is
obtained by taking independent variations with respect to the
fields defining the lagrangian functional. It turns out that
these lagrangians are singular ones. This implies that the
hamiltonian construction, via a Legendre transformation is
formulated on a constrained phase space. In all the cases we will
consider the constraints turn out to be primary constraints and
of the second class. The unconstrained phase space is equipped
with a Poisson bracket structure, however since there are second
class constraints we must obtain the Poisson bracket structure on
the constrained submanifold of the phase space. This Poisson
bracket is provided by the Dirac brackets \cite{Dirac}. It
satisfies all the properties of a Poisson bracket, in particular
the Jacobi identity. In this way starting from a lagrangian for
the system we can construct a Poisson bracket structure, together
with a hamiltonian functional. This approach was followed for the
KdV equation in \cite{Nutku,Kentwell}. It provides a geometrical
picture on phase space of the hamiltonian structure of the
integrable system. The other way to proceed is to find a
hamiltonian operator together with a hamiltonian functional.
Afterwards we may construct a Poisson bracket structure provided
the hamiltonian operator satisfies a differential restriction
\cite{Olver1} ensuring that the Jacobi identity is satisfied. In
this approach the set of allowed observables is only a subset of
the space of observables of the more general formulation in terms
of the constrained phase space approach.

We will obtain a pencil of Poisson bracket structures each of
them associated to a hamiltonian functional. In particular this
implies compatibility between some of the Poisson structures.

\section{The parametric coupled KdV system}We consider a coupled Korteweg-de Vries (KdV) system,formulated
in terms of two real differentiable functions $u(x,t)$ and
$v(x,t)$, given by the following partial differential equations:
\begin{eqnarray}& & u_t+uu_x+u_{xxx}+\lambda vv_x=0\\
& & v_t+u_xv+v_xu+v_{xxx}=0\end{eqnarray} where $\lambda$ is a
real parameter.

Here and in the sequel $u$ and $v$ belong to the real Schwartz
space defined by
\[C_\downarrow^\infty=\left\{w\in C^\infty( \mathbb{R})/\lim_{x\rightarrow \pm\infty} x^p\frac{\partial^q}{\partial
x^q}w=0;p,q\geq0 \right\}.\]

By a redefinition of $v$ given by
$v\rightarrow\frac{v}{\sqrt{|\lambda|}}$ we may reduce the values
of $\lambda>0$ to be $+1$ and $\lambda<0$ to be $-1$. The systems
for $\lambda=+1,\lambda=-1$ and $\lambda=0$ are not equivalent.
The $\lambda=-1$ case corresponds to the complexification of KdV
equation.

The case $\lambda=+1$ corresponds to two decoupled KdV equations.

The system (1),(2) for $\lambda=-1$ describes a two-layer liquid
model studied in references \cite{Gear,Lou,Brazhnyi}. It is a
very interesting evolution system. It is known to have solutions
developing singularities on a finite time \cite{Bona}. Also, a
class of solitonic solutions was reported in \cite{Yang} via the
Hirota approach \cite{Hirota1}.

The system (1),(2) for $\lambda=0$ corresponds to the ninth
Hirota-Satsuma \cite{Hirota} coupled KdV system given in
\cite{Sakovich} (for the particular value of $k=0$) (see also
\cite{Karasu}) and is also included in the interesting study
which relates integrable hierarchies with polynomial Lie algebras
\cite{Casati}.

(1),(2) is equivalent to the $ \mathcal{Z}_2^\lambda$-KdV equation
introduced in \cite{Zuo}. It was also considered from a different
point of view in \cite{Adrian4}.

A B\"{a}cklund transformation, the permutability theorem, the
Gardner transformation as well as the Gardner equations for the
coupled KdV system (1), (2), were obtained in \cite{Adrian5}. Also
a class of multisolitonic solutions and a class of periodic
solutions were found in \cite{Adrian5}.

\section{Poisson structures}In this and in the following section we will show that there exists four basic hamiltonians and four associated basic Poisson
structures for the coupled KdV system we are considering. We will
use the method of Dirac for constrained systems to deduce them.
The hamiltonian as defined in quantum physics must be a
selfadjoint operator conjugate to the time, hence our four
hamiltonians will be four real functionals in terms of the real
fields $w(x,t)$ and $y(x,t)$. We start our construction by
considering the lagrangian $L_1=\int_0^T
dt\int_{-\infty}^{+\infty} dx\,\mathcal{L}_1$,
\[\mathcal{L}_1=-\frac{1}{2}w_xw_t-\frac{1}{6}{w_x}^3+\frac{1}{2}{w_{xx}}^2-\frac{\lambda}{2}w_x{y_x}^2-\frac{\lambda}{2}y_xy_t+
\frac{\lambda}{2}{y_{xx}}^2\] for $\lambda\neq0,$ where
\begin{eqnarray*} u(x,t)&=&w_x(x,t)\\ v(x,t)&=&y_x(x,t).\end{eqnarray*}
By taking independent variations of $L_1$ with respect to $w$ and
to $y$ we obtain the field equations
\[\frac{\delta L_1}{\delta w}=0\hspace{4mm},\hspace{4mm}\frac{\delta L_1}{\delta y}=0,\]
which are the same as equations (1),(2).

We now introduce a second lagrangian $L_2=\int_0^T
dt\int_{-\infty}^{+\infty} dx \, \mathcal{L}_2$ where
\[\mathcal{L}_2=-\frac{1}{2}w_xy_t-\frac{1}{2}w_ty_x-\frac{1}{2}w_x^2y_x-y_xw_{xxx}-\frac{\lambda}{6}y_x^3\]
for any $\lambda$.

By taking independent variations of $L_2$ with respect to $w$ and
$y$ we obtain the same field equations.

We will now construct the hamiltonian structure associated to
each of these lagrangians. We start by considering the lagrangian
$L_1$. We introduce the conjugate momenta associated to $w$ and
$y$, we denote them $p$ and $q$ respectively, we have
\[p=\frac{\delta \mathcal{L}_1}{\delta
w_t}=-\frac{1}{2}w_x\hspace{5mm},\hspace{5mm} q=\frac{\delta
\mathcal{L}_1}{\delta y_t}=-\frac{\lambda}{2}y_x.\] We define
\[\phi_1\equiv
p+\frac{1}{2}w_x\hspace{5mm},\hspace{5mm}\phi_2=q+\frac{\lambda}{2}y_x.\]
$\phi_1$ and $\phi_2$ do not have any $w_t$ nor any $y_t$
dependence, hence $\phi_1=\phi_2=0$, they are constraints on the
phase space. It turns out that these are the only constraints on
the phase space. They are second class contraints.

The hamiltonian may be obtain directly from $\mathcal{L}_1$ by
performing a Legendre transformation,
\[\mathcal{H}_1=pw_t+qy_t-\mathcal{L}_1.\]
We obtain
\[\mathcal{H}_1=\frac{1}{6}w_x^3-\frac{1}{2}w_{xx}^2+\frac{\lambda}{2}w_xy_x^2-\frac{\lambda}{2}y_{xx}^2\]
and the corresponding hamiltonian $H_1=\int_{-\infty}^{+\infty}
dx\,\mathcal{H}_1.$

We introduce a Poisson structure on the phase space defined by
\begin{eqnarray*}\left\{w(x),p(\hat{x})\right\}_{PB}&=&\delta(x-\hat{x})\\\left\{y(x),q(\hat{x})\right\}_{PB}&=
&\delta(x-\hat{x})     \end{eqnarray*} with all other brackets
between these variables being zero.

Since we have a constrained phase space we must introduce the
Dirac brackets corresponding to a Lie bracket structure on the
constrained submanifold of phase space. The Dirac brackets
between two functionals $F$ and $G$ on phase space is defined as
\beq
\left\{F,G\right\}_{DB}=\left\{F,G\right\}_{PB}-{\left\langle{\left\langle\left\{F,\phi_i(x^\prime)\right\}_
{PB}\mathbb{C}_{ij}(x^\prime,x^{\prime\prime})\left\{\phi_j(x^{\prime\prime}),G\right\}_
{PB}\right\rangle}_{x^\prime}\right\rangle}_{x^{\prime\prime}}
\eeq where $<>_{x^\prime}$ denotes integration on $x^\prime$ from
$-\infty$ to $+\infty$. The indices $i,j=1,2$ and the $
\mathbb{C}_{ij}(x^\prime,x^{\prime\prime})$ are the components of
the inverse of the matrix whose components are
$\left\{\phi_i(x^\prime),\phi_j(x^{\prime\prime}\right\}_{PB}$.

This matrix becomes
\[\left[\begin{array}{cc}\partial_{x^\prime}\delta(x^\prime-x^{\prime\prime})&0
\\0&\lambda\partial_{x^\prime}\delta(x^\prime-x^{\prime\prime})\end{array}\right]\] and its inverse, satisfying
\begin{eqnarray*}&&{\left\langle\left[\begin{array}{cc}\partial_x\delta(x-x^{\prime\prime})&0
\\0&-\partial_x\delta(x-x^{\prime\prime})\end{array}\right]
\left[\begin{array}{cc}\mathbb{C}_{11}(x^{\prime\prime},\hat{x})&\mathbb{C}_{12}(x^{\prime\prime},\hat{x})
\\\mathbb{C}_{21}(x^{\prime\prime},\hat{x})&\mathbb{C}_{22}(x^{\prime\prime},\hat{x})\end{array}\right]
\right\rangle}_{x^{\prime\prime}}=\\
&=&\left[\begin{array}{cc}\delta(x-\hat{x})&0
\\0&\delta(x-\hat{x})\end{array}\right]\end{eqnarray*} is given by
\begin{eqnarray*}\left[\mathbb{C}_{ij}(x^\prime,x^{\prime\prime})\right]=\left[\begin{array}{cc}\int^{x^\prime}\delta(s-x^{\prime\prime})ds&0
\\0&\frac{1}{\lambda}\int^{x^\prime}\delta(s-x^{\prime\prime})ds\end{array}\right].\end{eqnarray*}
It turns out, after some calculations, that
\begin{eqnarray*}&&\left\{u(x),u( \hat{x})\right\}_{DB}=-\partial_x\delta(x-\hat{x})\hspace{2mm},\hspace{2mm}\left\{v(x),v( \hat{x})\right\}_{DB}=
-\frac{1}{\lambda}\partial_x\delta(x-\hat{x})\\&&\left\{u(x),v(
\hat{x})\right\}_{DB}=0. \end{eqnarray*} We notice that this
Poisson bracket is not well defined for $\lambda=0$. We have
already assume $\lambda\neq0$.

 From them we obtain the Hamilton
equations, which are of course the same as (1),(2):
\beq \begin{array}{cc}u_t=\left\{u,H_1\right\}_{DB}=-uu_x-u_{xxx}-\lambda vv_x\\
 v_t=\left\{v,H_1\right\}_{DB}=-u_xv-v_xu-v_{xxx}.\end{array}\eeq
 Moreover, we may obtain directly the Dirac bracket of any two functionals $F(u,v)$ and $G(u,v)$ from (3)
 using the above bracket relations for $u$ and $v$. We notice that
 the observables $F$ and $G$ in (3) may be functionals of $w,y,p$
 and $q$, not only of $u$ and $v$. In this sense the phase space
 approach for singular lagrangians provides the most general space
 of observables. The same comment will be valid for the phase
 space construction using lagrangians $L_2$ and $ L_1^M,L_2^M$ in
 the following sections.

 We now consider the lagrangian $L_2$ and its associated hamiltonian structure. In this case we denote the conjugate momenta to $w$ and $y$ by $ \hat{p}$ and $\hat{q}$ respectively. We have
 \[\hat{p}=-\frac{1}{2}y_x\hspace{4mm},\hspace{4mm}\hat{q}=-\frac{1}{2}w_x.\]
 The constraints become in this case
 \[\widehat{\phi_1}=\hat{p}+\frac{1}{2}y_x=0\hspace{4mm},\hspace{4mm}\widehat{\phi_2}=\hat{q}+\frac{1}{2}w_x=0.\]
 The corresponding Poisson brackets between $\phi_i$ and $\phi_j,i,j=1,2,$ are given by
 \begin{eqnarray*}&&\left\{\widehat{\phi_1}(x),\widehat{\phi_1}(x^\prime)\right\}_{PB}=0\hspace{4mm},\hspace{4mm}
 \left\{\widehat{\phi_2}(x),\widehat{\phi_2}(x^\prime)\right\}_{PB}=0,\\&&\left\{\widehat{\phi_1}(x),
 \widehat{\phi_2}(x^\prime)\right\}_{PB}=
 \partial_x\delta(x-x^\prime). \end{eqnarray*}
 The corresponding construction of the Dirac brackets yields
  \begin{eqnarray*}&&\left\{u(x),u(\hat{x})\right\}_{DB}=0\hspace{4mm},\hspace{4mm}
 \left\{v(x),v(\hat{x})\right\}_{DB}=0,\\&&\left\{u(x),v(\hat{x})\right\}_{DB}=
 -\partial_x\delta(x-\hat{x}). \end{eqnarray*}
 The hamiltonian $H_2=\int_{-\infty}^{+\infty} dx\,\mathcal{H}_2$ is given
 by the hamiltonian density
 \[\mathcal{H}_2=\frac{1}{2}w_x^2y_x+y_xw_{xxx}+\frac{\lambda}{6}y_x^3.\]
 The Hamilton equations
 \[u_t(x)=\left\{u(x),H_2\right\}_{DB}\hspace{4mm},\hspace{4mm}v_t(x)=\left\{v(x),H_2\right\}_{DB}\] now using
 the corresponding
 Dirac brackets yields the same fields equations (1),(2) for any $\lambda$. We have thus constructed two hamiltonian functionals and associated
 Poisson bracket structures. These two hamiltonian structures arise directly from the basic lagrangians $L_1$ and
 $L_2$. We will now construct two additional hamiltonian structures by considering the Miura transformation.

 The hamiltonians $H_1$ and $H_2$, $H_1^M$ and $H_2^M$ in the following section, were
 presented in \cite{Zuo}.

 \section{The Miura transformation}We consider the Miura transformation
 \beq\begin{array}{ll}u=\mu_x-\frac{1}{6}\mu^2-\frac{\lambda}{6}\nu^2\\v=\nu_x-\frac{1}{3}\mu\nu.\end{array}\eeq
 The corresponding modified KdV system (MKdVS) is given by
 \beq\begin{array}{lll}\mu_t+\mu_{xxx}-\frac{1}{6}\mu^2\mu_x-\frac{\lambda}{6}\nu^2\mu_x-\frac{\lambda}{3}\mu\nu\nu_x=0
 \\ \\  \nu_t+\nu_{xxx}-\frac{1}{6}\mu^2\nu_x-\frac{\lambda}{6}\nu^2\nu_x-\frac{1}{3}\mu\nu\mu_x=0.\end{array}\eeq
 These equations may be obtained from two lagrangians, which we will denote $L_1^M=\int_0^T dt\int_{-\infty}^{+\infty} dx\,
 \mathcal{L}_1^M$ and $L_2^M=\int_0^T dt\int_{-\infty}^{+\infty} dx\, \mathcal{L}_2^M.$

 The lagrangian densities $\mathcal{L}_1^M$, formulated for $\lambda\neq0$, and $\mathcal{L}_2^M$, formulated for any $\lambda$,
  expressed in terms of $\sigma,\rho$ where $\mu=\sigma_x,
\nu=\rho_x$ are given by
  {\beq
\mathcal{L}_1^M=-\frac{1}{2}\sigma_t\sigma_x-\frac{\lambda}{2}\rho_t\rho_x-\frac{1}{2}\sigma_x\sigma_{xxx}-
\frac{\lambda}{2}\rho_x\rho_{xxx}
+\frac{1}{72}{\sigma_x}^4-\frac{\lambda^2}{72}{\rho_x}^4+\frac{\lambda}{12}\rho_x^2\sigma_x^2
\eeq } and {\beq
\mathcal{L}_2^M=-\frac{1}{2}\sigma_t\rho_x-\frac{1}{2}\sigma_x\rho_t-\sigma_{xxx}\rho_x+\frac{1}{18}{\sigma_x}^3\rho_x+
\frac{\lambda}{18}{\rho_x}^3\sigma_x \eeq } respectively.

We will now construct the hamiltonian structure associated to
$\mathcal{L}_1^M.$

We denote by $\alpha$ and $\beta$ the conjugate momenta
associated to $\sigma$ and $\rho$ respectively. We have
\[\alpha=\frac{\delta \mathcal{L}_1^M}{\delta \sigma_t}=-\frac{1}{2}\sigma_x\hspace{5mm},\hspace{5mm}
\beta=\frac{\delta \mathcal{L}_1^M}{\delta
\rho_t}=-\frac{\lambda}{2}\rho_x.\] These are constraints on the
phase space.

The hamiltonian $H_1^M$ corresponding to this lagrangian density
$ \mathcal{L}_1^M$ is given by
\begin{eqnarray*}&&\mathcal{H}_1^M=v^2-u^2\\&&H_1^M=\int_{-\infty}^{+\infty}\mathcal{H}_1^M\,dx \end{eqnarray*}
where $u$ and $v$ are given in terms of $\mu$ and $\nu$ by the Miura transformation.

The construction of the Dirac brackets follows in the usual way.
We end up with the following Poisson structure on the constrained
submanifold,
\begin{eqnarray*}\left\{\mu(x),\mu(\hat{x})\right\}_{DB}&=&-\partial_x\delta(x,\hat{x})\\
\left\{\nu(x),\nu(\hat{x})\right\}_{DB}&=&-\frac{1}{\lambda}\partial_x\delta(x,\hat{x})\\
\left\{\mu(x),\nu(\hat{x})\right\}_{DB}&=&0.\end{eqnarray*}

From these Poisson bracket structure we obtain for the original
$u$ and $v$ fields
\begin{eqnarray*}\left\{u(x),u(\hat{x})\right\}_{DB}&=&\partial_{xxx}\delta(x,\hat{x})+\frac{1}{3}u_x\delta(x,\hat{x})+
\frac{2}{3}u\partial_x\delta(x,\hat{x})\\\left\{v(x),v(\hat{x})\right\}_{DB}&=&\frac{1}{\lambda}
\partial_{xxx}\delta(x,\hat{x})+\frac{1}{3\lambda}u_x\delta(x,\hat{x})+
\frac{2}{3\lambda}u\partial_x\delta(x,\hat{x})\\\left\{u(x),v(\hat{x})\right\}_{DB}&=&\frac{1}{3}v_x\delta(x,\hat{x})+
\frac{2}{3}v\partial_x\delta(x,\hat{x})\end{eqnarray*} which
defines the Poisson structure on the original fields inherited
from the Poisson structure on the constrained submanifold on the
phase space associated to the modified KdV system. This Poisson
bracket is not well defined for $\lambda\neq0$. We have already
assumed $\lambda\neq0.$

From the Dirac brackets of $u$ and $v$ we may obtain directly the
hamiltonian field equations
\beq \begin{array}{cc}u_t=\left\{u,H_1^M\right\}_{DB}=-uu_x-u_{xxx}-\lambda vv_x\\ \\
 v_t=\left\{v,H_1^M\right\}_{DB}=-v_{xxx}-{(uv)}_x\end{array}\eeq which, as it should be, coincide with system (1),(2).

 We have then obtained the Poisson structure associated to the hamiltonian $H_1^M.$

 We now proceed to obtain a second Poisson structure starting from the Lagrangian $ \mathcal{L}_2^M.$

 The hamiltonian obtained via a Legendre transformation is given by
 $H_2^M=\int_{-\infty}^{+\infty}(-uv)\,dx$ where $u$ and $v$ are functions of $\mu$ and $\nu$ according to the Miura transformation.
 We use as before $\mu=\sigma_x,\nu=\rho_x.$

 We denote by $\hat{ \alpha}$ and $ \hat{\beta}$ the conjugate
 momenta associated to $\sigma$ and $\rho$ respectively.

 The constraints on phase space become now
 \begin{eqnarray*}\hat{\alpha} &=&-\frac{1}{2}\rho_x\\ \hat{\beta} &=&-\frac{1}{2}\sigma_x. \end{eqnarray*}

 The Dirac brackets are
 \begin{eqnarray*}\left\{\mu(x),\mu(\hat{x})\right\}_{DB}&=&0\\\left\{\nu(x),\nu(\hat{x})\right\}_{DB}&=&0
 \\\left\{\mu(x),\nu(\hat{x})\right\}_{DB}&=&-\partial_x\delta(x,\hat{x}).\end{eqnarray*}

We then obtain, for any $\lambda$,
\begin{eqnarray*}\left\{u(x),u(\hat{x})\right\}_{DB}&=&\frac{\lambda}{3}v_x\delta(x,\hat{x})+\frac{2\lambda}{3}v\partial_x\delta(x,\hat{x})
\\\left\{v(x),v(\hat{x})\right\}_{DB}&=&\frac{1}{3}
v_x\delta(x,\hat{x})+\frac{2}{3}v\partial_x\delta(x,\hat{x})\\\left\{u(x),v(\hat{x})\right\}_{DB}&=
&\partial_{xxx}\delta(x,\hat{x})+\frac{1}{3}u_x\delta(x,\hat{x})
+\frac{2}{3}u\partial_x\delta(x,\hat{x}).\end{eqnarray*}

This is the Poisson bracket structure inherited from the second
Poisson structure on the modified phase space. One may directly
verify that the corresponding Hamilton equations exactly coincide
with equations (1),(2). We have then constructed four basic
lagrangians and associated hamiltonian functionals together with
four basic Poisson structures.
\section{Two pencils of Poisson structures for the coupled system} We now construct a parametric
lagrangian density $ \mathcal{L}_k$, where $k$ is a real
parameter, associated to the two basic lagrangians $L_1$ and
$L_2$.

 We define the lagrangian density
\[\mathcal{L}_k=k\mathcal{L}_1+\left(1-k\right)\mathcal{L}_2.\]
The field equations obtained from this lagrangian density are
equivalent to (1) and (2) in the following cases: If $\lambda<0$
for any $k$. If $\lambda=0,$ for $k\neq1.$ If $\lambda>0$ for
$k\neq\frac{1}{1+\sqrt{\lambda}}$ and
$k\neq\frac{1}{1-\sqrt{\lambda}}$. From now on we will excluded
this particular values of $k$. The corresponding hamiltonian
density is given by
\[\mathcal{L}_k=pw_t+qy_t-\mathcal{L}_k=k\mathcal{H}_1+\left(1-k\right)\mathcal{H}_2\] and the primary constraints by
\begin{eqnarray}\phi_1&\equiv &\frac{k}{2}w_x+\frac{\left(1-k\right)}{2}y_x+p=0\\\phi_2&\equiv & \frac{\lambda k}{2}y_x+
\frac{\left(1-k\right)}{2}w_x+q=0.\end{eqnarray}

These are the only constraints on phase space, they are second
class ones.

 The Poisson brackets on the unconstrained phase space are
\begin{eqnarray*}\left\{\phi_1(x),\phi_1(\hat{x}) \right\}_{PB}&=&k\partial_x\delta(x,\hat{x})
\\ \left\{\phi_2(x),\phi_2(\hat{x}) \right\}_{PB}&=&\lambda k\partial_x\delta(x,\hat{x})\\
\left\{\phi_1(x),\phi_2(\hat{x})
\right\}_{PB}&=&\left(1-k\right)\partial_x\delta(x,\hat{x}).
\end{eqnarray*}

We will denote by $\left\{\right\}_{DB}^k$ the Dirac bracket
corresponding to the parameter $k$.

The Dirac brackets are then given by
\begin{eqnarray*}\left\{u(x),u(\hat{x}) \right\}_{DB}^k&=&\frac{\lambda k}{-\lambda k^2+{(1-k)}^2}\partial_x\delta(x,\hat{x})\\
\left\{v(x),v(\hat{x}) \right\}_{DB}^k&=&\frac{k}{-\lambda k^2+{(1-k)}^2}\partial_x\delta(x,\hat{x})\\
\left\{u(x),v(\hat{x}) \right\}_{DB}^k&=&\frac{1-k}{-\lambda
k^2+{(1-k)}^2}\left(-\partial_x\delta\left(x,\hat{x}\right)\right),
\end{eqnarray*} where the denominator is different from zero for the values of $k$ we are considering. They define the Poisson structure for the
hamiltonian $H_k=\int_{-\infty}^{+\infty}\,\mathcal{H}_kdx$.

The associated Hamilton equations coincide with the coupled
equations (1),(2). It is interesting to notice that the above
Poisson structure is a linear combination of the Dirac brackets
introduced associated to hamiltonians $H_1$ and $H_2$. In the
present notation $H_2$ corresponds to $k=0.$

We then have
\[\left\{F,G\right\}_{DB}^k=\frac{-\lambda k}{-\lambda k^2+{(1-k)}^2}\left\{F,G\right\}_{DB}^1+\frac{1-k}{-\lambda k^2+
{(1-k)}^2}\left\{F,G\right\}_{DB}^0\] where $F,G$ are any
functionals of $u$ and $v$. In particular for any $\lambda$
different from one and zero, and $k=\frac{1}{1-\lambda}$, we
obtain
\[\left\{F,G\right\}_{DB}^k=\left\{F,G\right\}_{DB}^1+\left\{F,G\right\}_{DB}^0.\]
Consequently, the two basic Poisson brackets for every
$\lambda\neq0,1$ are then compatible.

We also notice that for any $k$ and $\lambda=-1$, using the above
Poisson bracket structure, one gets
\begin{eqnarray}& &\left\{u(x)+iv(x),u(\hat{x})-iv(\hat{x} )\right\}_{DB}^k=0,\\&&
\left\{u(x)+iv(x),u(\hat{x})+iv(\hat{x} )\right\}_{DB}^k=
-\frac{2}{k^2+{(1-k)}^2}\partial_x\delta(x,\hat{x}).
\end{eqnarray}

We emphasize that only (13) arises from the complexification of
the corresponding Poisson structure for real KdV. The relation
(12) follows in our approach from first principles. It is not
imposed by hand.  The existence of a local real hamiltonian $H_k$
for each $k$ is a non-trivial feature of the system (1),(2) and
is not an algebraic consequence of the complexification of the
real KdV equation.

We may now consider the case $\lambda=0$. The Poisson bracket for
any $k\neq1$ becomes
\beq\left\{F,G\right\}_{DB}^k=\frac{k}{2{(1-k)}^2}\left\{F,G\right\}_{DB}^{\frac{1}{2}}+\frac{1-2k}{{(1-k)}^2}\left\{F,G\right\}_{DB}^0\eeq
in particular for $k=\frac{5}{2}$ the two coefficients are equal,
hence the Poisson brackets for $k=\frac{1}{2}$ and $k=0$ are
compatible.

We have thus constructed a pencil of Poisson structures, each of
them with an associated local real hamiltonian
$H_k=\int_{-\infty}^{+\infty}\,\mathcal{H}_k$.

We now construct, as we have already done with $ \mathcal{L}_1$
and $ \mathcal{L}_2$, a parametric lagrangian density $
\mathcal{L}_k^M=k\mathcal{L}_1^M+(1-k)\mathcal{L}_2^M$. The
associated hamiltonian density is given by $
\mathcal{H}_k^M=k\mathcal{H}_1^M+(1-k)\mathcal{H}_2^M$ in terms
of the other two basic lagrangian densities. The constraints on
phase space are given by
\begin{eqnarray*}\phi_1&\equiv&\alpha+\frac{k}{2}\sigma_x+\frac{(1-k)}{2}\rho_x=0\\
\phi_2&\equiv&\beta\frac{\lambda
k}{2}\rho_x+\frac{(1-k)}{2}\sigma_x=0
\end{eqnarray*} which are second class constraints and the only contraints on the phase space. The Poisson
brackets on the unconstrained phase space are
\begin{eqnarray*}\left\{\phi_1(x),\phi_1(\hat{x}) \right\}_{PB}&=&k\partial_x\delta(x,\hat{x})
\\ \left\{\phi_2(x),\phi_2(\hat{x}) \right\}_{PB}&=&\lambda k\partial_x\delta(x,\hat{x})\\
\left\{\phi_1(x),\phi_2(\hat{x})
\right\}_{PB}&=&\left(1-k\right)\partial_x\delta(x,\hat{x}).
\end{eqnarray*} and the Dirac brackets are then given by
\beq\begin{array}{lllll}\left\{u(x),u(\hat{x})
\right\}_{DB}^k&=&-\frac{\lambda k}{-\lambda k^2+{(1-k)}^2}
\left(\partial_{xxx}\delta(x,\hat{x})+\frac{1}{3}u_x\delta(x,\hat{x})+
\frac{2}{3}u\partial_x\delta(x,\hat{x})\right)- \\
&-&\frac{1-k}{-\lambda
k^2+{(1-k)}^2}\left(\frac{1}{3}v_x\delta(x,\hat{x})
+\frac{2}{3}v\partial_x\delta(x,\hat{x})\right) \\
\left\{u(x),v(\hat{x}) \right\}_{DB}^k&=&\frac{(1-k)}{-\lambda
k^2+{(1-k)}^2}\left(\partial_{xxx}\delta(x,\hat{x})+\frac{1}{3}u_x\delta(x,\hat{x})+
\frac{2}{3}u\partial_x\delta(x,\hat{x})\right)+ \\ &-&\frac{\lambda k}{-\lambda k^2+{(1-k)}^2}\left
(\frac{1}{3}v_x\delta(x,\hat{x})+\frac{2}{3}v\partial_x\delta(x,\hat{x})\right)\\
\left\{v(x),v(\hat{x}) \right\}_{DB}^k&=&-\frac{k}{-\lambda
k^2+{(1-k)}^2}\left(\partial_{xxx}\delta\left(x,\hat{x}\right)+\frac{1}{3}u_x\delta(x,\hat{x})
+\frac{2}{3}u\partial_x\delta(x,\hat{x})\right)+\\
&+&\frac{(1-k)}{-\lambda
k^2+{(1-k)}^2}\left(\frac{1}{3}v_x\delta(x,\hat{x})+\frac{2}{3}v\partial_x\delta(x,\hat{x})\right).\end{array}
\eeq It follows from the construction that the Hamilton equations
in terms of the corresponding Poisson structure,
\begin{eqnarray*}u_t=\left\{u(x),H_k^M)\right\}_{DB}^k\hspace{2mm},\hspace{2mm}v_t=\left\{v(x),H_k^M)
\right\}_{DB}^k \end{eqnarray*} are equivalent to the coupled KdV
system (1),(2).

As in the previous case the pencil of Poisson structures can be
rewritten in terms of the basic Poisson structures which
corresponds to $k=1$ and $k=0$ in (14):

\[\left\{F,G\right\}_{DB}^k=\frac{-\lambda k}{-\lambda k^2+{(1-k)}^2}\left\{F,G\right\}_{DB}^1+\frac{1-k}{-\lambda k^2+
{(1-k)}^2}\left\{F,G\right\}_{DB}^0.\]

We notice that this decomposition is the same as in previous
case, however the basic Poisson structure are different.

In particular for $k=\frac{1}{1-\lambda}$, $\lambda\neq0,1$, the
$\left\{,\right\}_{DB}^k$ is the sum of the
$\left\{,\right\}_{DB}^1$ and $\left\{,\right\}_{DB}^0$ basic
Poisson structures. For $\lambda=0$ and $k\neq1$ the same
relation (14) holds for the Poisson bracket we are now
considering. These are then compatible Poisson structures.

We notice that by construction $\phi_1$ and $\phi_2$ as well as
any functional of them, in all the cases we have considered, are
Casimirs of the Poisson structure defined in terms of the Dirac
brackets. In fact,
\begin{eqnarray*}\left\{F,\phi_1)  \right\}_{DB}&=&0\\\left\{F,\phi_2)  \right\}_{DB}&=&0\end{eqnarray*} for any functional $F$ on phase
space. This is a general property of the Dirac bracket.

It is a non-trivial feature that for each real $k$, the parameter
of the pencil of Poisson structures, there are hamiltonians $H_k$
and $H_k^M$ which give rise to the coupled KdV system when the
corresponding Poisson structure is used.

\section{Conclusions}We obtained the full hamiltonian structure
for a coupled parametric KdV system. We started from four basic
singular lagrangians. The associated hamiltonian formulation on
phase space is restricted by second class constraints. The
Poisson structure on the constrained variety of phase space was
obtained using the Dirac approach. The Dirac brackets on the
constrained phase space yields the most general structure of
observables. A subset of them are functionals of the original
fields $u(x,t),v(x,t)$ of the coupled KdV system. We then
constructed two pencils of Poisson brackets each of them with an
associated parametric hamiltonian in terms of the same parameter
of each pencil.

Each pencil of Poisson brackets is obtained from two compatible
Poisson brackets of the same dimension. Consequently it is not
possible to construct a hierarchy of higher dimensional
hamiltonians from them. However the two pencils of Poisson
brackets are of different dimensions, hence one may construct a
hierarchy of higher order hamiltonians as in the KdV case.

\newpage

\textbf{Acknowledgments} A. R. and A. S. are partially supported
by Projects Fondecyt 1121103 and Mecesup ANT398 (Chile).

We thank Professors P. Casati and S. Krivonos for fruitful
discussions.


\begin{thebibliography}{}
\bibitem{Hirota}{R. Hirota and J. Satsuma, Phys. Lett. 85A, number 8,9 407-408 (1981).}
\bibitem{Gear}{J. A. Gear and R. Grimshaw, Stud. Appl. Math. 70, 235 (1984); J. A. Gear, Stud. Appl. Math. 72, 95
(1985).}
\bibitem{Lou}{S. Y. Lou, B. Tong, H. C. Hu and X. Y. Tang, J. Phys. A: Math. Gen. 39,
513-527 (2006).}
\bibitem{Karasu}{A. K. Karasu, J. Math. Phys. 38 (7), 3616-3622 (1997).}
\bibitem{Sakovich}{S. Yu. Sakovich, J. Nonlin. Math. Phys. 6, N 3 255-262 (1999).}
\bibitem{Casati}{P. Casati and G. Ortenzi, J. Geom. and Phys. 56, 418-449 (2006).}
\bibitem{Kupershmidt}{B. A. Kupershmidt 102A, N 5,6 213-215 (1984).}
\bibitem{Mathieu1}{P. Mathieu, J. Math. Phys. 29, 2499 (1988).}
\bibitem{Mathieu2}{P. Labelle and P. Mathieu, J. Math. Phys. 32, 923 (1991).}
\bibitem{Krivonos1}{S. Belucci, E. Ivanov and S. Krivonos, J. Math. Phys. 34, 3087 (1993).}
\bibitem{Delduc}{F. Delduc and E. Ivanov, Phys. Lett. B 309, 312-319 (1993).}
\bibitem{Krivonos2}{F. Delduc, E. Ivanov and S. Krivonos, J. Math. Phys. 37, 1356 (1996).}
\bibitem{Popowicz}{Z. Popowicz, Phys. Lett. B 459, 150-158 (1999).}
\bibitem{Olver}{P. J. Olver and V. V. Sokolov, Commun. Math. Phys. 193(2), 245-268 (1998).}
\bibitem{Restuccia}{A. Restuccia and A. Sotomayor, J. Math. Phys. 54, 113510 (2013).}
\bibitem{Dirac}{P. A. M. Dirac, ``Lectures on Quantum Mechanics", Belfer Graduate School Monograph
Series No.2, Yeshiva University, New York, (1964).}
\bibitem{Nutku}{Y. Nutku, J. Math. Phys. 25 (6), June (1984).}
\bibitem{Kentwell}{G. W. Kentwell, J. Math. Phys. 29, 46 (1988).}
\bibitem{Olver1}{P. J. Olver, Y. Nutku, J. Math. Phys. 29 (7) July 1988.}
\bibitem{Brazhnyi}{V. A. Brazhnyi and V. V. Konotop, Phys. Rev. E 72,
026616 (2005).}
\bibitem{Bona}{J. L. Bona, S. Vento and F. B. Weissler, Discrete and Continuous Dynamical Systems 33, V 11-12, 4811 (2013).}
\bibitem{Yang}{J. R. Yang and J. J. Mao, Commun. Theor. Phys. 49, 22-26 (2008).}
\bibitem{Hirota1}{R. Hirota, Phys. Rev. Lett. 27, number 18, 1192-1194 (1971).}
\bibitem{Zuo}{D. Zuo, arxiv: 1403.0027 v1 [math-ph].}
\bibitem{Adrian4}{L. C. Vega, A. Restuccia and A. Sotomayor, Contribution to the Proceedings of the
2nd International Conference on Mathematical Modeling in Physical
Sciences 2013, Journal of Physics: Conference Series 490 (2014)
012024.}
\bibitem{Adrian5}{L. C. Vega, A. Restuccia and A. Sotomayor,  arXiv:1407.7743 v1 [math-ph].}

\end{thebibliography}
\end{document}